\documentclass[12pt,a4paper,dvips]{article}
\usepackage{cite,mcite}
\usepackage{graphicx}
\graphicspath{{/l3/paper/example/figs/}}
\newlength{\capindent}
\newlength{\capwidth}
\newlength{\figwidth}
\newcommand{\icaption}[2][!*!,!]{\hspace*{\capindent}%
  \begin{minipage}{\capwidth}
    \ifthenelse{\equal{#1}{!*!,!}}%
      {\caption{#2}}%
      {\caption[#1]{#2}}
  \end{minipage}}
\newcommand {\Be}{\begin{equation}}
\newcommand {\Ee}{\end{equation}}

\newcommand {\eqref}[1]{equation~(\ref{#1})}

\newcommand {\Figref}[1]{Figure~\ref{fig:#1}}

\renewcommand{\thefootnote}{\fnsymbol{footnote}}

\begin{document}

\begin{titlepage}
\begin{flushright} 
astro-ph/9907078 \\
ETHZ-IPP PR-99-04 \\
July  6, 1999 
\end{flushright}

\vspace*{3.0cm}

\begin{center} {\Large \bf
       Simulation of Multi-muon Events from EAS\\
\vspace*{0.15cm}
       at Shallow Depths Underground}

\vspace*{2.0cm}
  {\Large
  Dimitri Bourilkov\footnote{\tt e-mail: Dimitri.Bourilkov@cern.ch}
  }

\vspace*{1.0cm}
  Institute for Particle Physics (IPP), ETH Z\"urich, \\
  CH-8093 Z\"urich, Switzerland
\vspace*{4.3cm}
\end{center}

%
%
%
\begin{abstract}
The Monte Carlo program ARROW, based on GEANT and using GHEISHA at energies
below 30 GeV, is developed for simulation of the hadron and muon
components of extensive air showers with primary energy
$\rm 10^{12}-10^{17}$ eV.
Calculations of the characteristics of multi-muon events as observed
underground by the LEP detectors are presented and their dependence on
the primary cosmic ray composition and some basic assumptions of the
hadronic interaction model is discussed. 
\end{abstract}

\vspace*{1.0cm}
\begin{center}
{\it Talk presented at the First Arctic Workshop on Cosmic\\
      Ray Muons, Sodankyl\"a, Finland, April 24-29, 1999}
\end{center}

\end{titlepage}

\renewcommand{\thefootnote}{\arabic{footnote}}
\setcounter{footnote}{0}
%
%
\section*{Introduction}

The muon component of extensive air showers (EAS), due to the long muon range
in the Earth's atmosphere, carries a wealth of information about the
shower development. Study of multi-muon events gives an insight into the
primary cosmic ray composition and the physics of high energy hadronic
interactions.
The LEP detectors, situated between 30 and 150 m underground, offer
interesting possibilities to detect and study such
events~\cite{l3note_1676,l3note_1977,cosmolep_1}, which are
complimentary to the data collected in traditional cosmic ray
experiments.
The hadron component of EAS is absorbed, while the muon component is
detected with low threshold (typically, if we exclude access shafts, between
15 and 75 GeV) and high momentum and spatial resolution by the sophisticated
tracking systems of the LEP detectors. The multi-muon event rate is high enough
to make studies of the knee region possible with one year of data taking.

The interpretation of the measurements at the Lake of Geneva level
needs a detailed simulation of the interactions of the primary cosmic
ray particles with the air nuclei in the upper atmosphere and the
consequent shower development through air and the rock overburden of
the detectors. Many models are in use, and all of them rely on extrapolations
from the existing accelerator data at lower energies. This fact, together with
the inherent ambiguity between the inelasticity in the first interactions
and the mass composition of the primary cosmic rays, introduces a model
dependence in the extraction of quantities of physics interest from the
measurements and makes the use of many observables and different models
desirable.

%
%
\section*{The Method}

In~\cite{DB} a method for simulation of the hadron component of EAS and
the code ARROW are developed. They are based on the well known MC
simulation programs GEANT~\cite{mygeant} and GHEISHA~\cite{mygheisha},
widely in use in high energy
physics. GHEISHA simulates in great detail the nuclear interactions
up to $\sim 30-50$ GeV. For higher energies up to $\rm 10^{17}$ eV
a model based on extrapolations from existing accelerator data on
total and elastic cross sections, charged particle multiplicities and
leading particle spectra is developed. This allows to study the
dependence of key characteristics of EAS on the basic assumptions
of the hadronic interaction model. For primary nuclei the superposition
model is used. As GEANT defines {\em only} volumes with constant density,
the Earth's atmosphere is modeled with 30 layers, giving a suitable
description of the real density profile. The modeling of the overburden
and the detector volumes is straightforward.

In this contribution I extend the code ARROW to study the muon component
of EAS. Here one can take advantage of the  well established tracking 
capabilities of GEANT. The muonic interactions are simulated up to 10 TeV:
      \begin{itemize}
       \item  decay in flight
       \item  ionisation and $\delta$-ray production
       \item  multiple scattering
       \item  bremsstrahlung
       \item  direct ($\rm e^+e^-$) pair production
       \item  nuclear interactions
      \end{itemize}
making GEANT useful for cosmic ray studies.

%
%
\section*{Results and discussion}

As a case study
simulations for the L3 setup are performed for vertical showers initiated
by protons or iron with primary energies 1, 10, 100 TeV, 1, 10, 100 PeV.
The altitude is 449 m above sea level, corresponding to $\sim$~980 g/cm$\rm ^2$.
All muons which reach the level of the muon chambers of L3 after passing
the earth overburden (28.75 m of molasse),
the magnet coil and the return yoke are retained for
further analysis. The minimum energy at this stage is required to be 2 GeV
in order to have good tracks in the chambers.
The variation of the muon flux is found to be:
\begin{itemize}
 \item below 10 \% if we vary $\rm \bar x_{leading}$ from 0.25 to 0.30
 \item increase by 20 \% for protons at 1 PeV if we use a fast increase in
       charged particle multiplicity~\cite{DB}.
\end{itemize}
The results presented further use the parametrization from~\cite{chmul},
which predicts a slow increase of the charged multiplicity and can be
considered as a safe lower limit for the expected muon multiplicity.

The total number of muons in the shower above given threshold is
parametrized as:
$$\rm N_{\mu} = A\cdot P_1 \cdot (E/A)^{P_2} \cdot (1-\frac{P_3 \cdot E^{thr}_{\mu}}{E/A})^{10}$$
where A is the atomic number and E the total energy in GeV of the primary,
$\rm E^{thr}_{\mu}$ is the muon threshold energy, and $\rm P_1,\ P_2,\ P_3$
are free parameters. The results are summarized in Table~\ref{nmuon}.

\begin{table}
 \renewcommand{\arraystretch}{1.2}
  \begin{center}
    \begin{tabular}{|c|ccc|}
\hline
~~Primary~~&~~~P$_1$~~&~~~P$_2$~~&~~~P$_3$~~\\
\hline
proton     &  0.0129  &  0.798   &    0.00  \\
iron       &  0.0190  &  0.766   &    1.04  \\
\hline
    \end{tabular}
  \end{center}
  \caption{
    Results of the fit to the ARROW simulations for the total number of
    muons in EAS initiated by protons or iron with energies from 1 TeV to
    100 PeV
    at L3 level. For protons the threshold factor is set to zero.}
  \label{nmuon}
\end{table}

The muon density is fitted with the Greisen function
($\rm P_1 = N_{\mu},\ P_2 = R_0$):
$$\rm \rho_{\mu}(r) = 0.2575*P_1\cdot (\frac{1}{P_2})^{1.25}\cdot r^{-0.75}\cdot (1+\frac{r}{P_2})^{-2.5}$$
The results are summarized in Table~\ref{muden}. The two independent fits
give similar results for the total number of muons, showing an overall good
description of the muon density by the Greisen function. For energies below
$\sim$~100 TeV the $\rm \chi^2$ of the fits is rising, indicating deviations
of the muon density closer to threshold from this simple functional form.
 One can observe
that the parameter $\rm R_0$ is shrinking for higher energies, reflecting
a build-up of very high muon densities near the core for more energetic showers.
The energy dependence of $\rm R_0$ is parametrized with a third order
polynomial in $\rm \log(E)$.

\begin{table}
 \renewcommand{\arraystretch}{1.2}
  \begin{center}
    \begin{tabular}{|c||r|r||r|r|}
\hline
~~Energy~~&~~~N$_{\mu}$~~&~~R$_0$~~~&~~~N$_{\mu}$~~&~~R$_0$~~~\\
   PeV    &              &~~~~m~~~&                &~~~m~~~\\
\hline
          & \multicolumn{2}{c||}{proton}& \multicolumn{2}{c|}{iron}\\
\hline 
  0.01    &        22    &  127     &      18    &  493      \\
   0.1    &       115    &  127     &     270    &  245      \\
     1    &       750    &  101     &    1800    &  124      \\
    10    &      5200    &   86     &   10800    &  109      \\
   100    &     30700    &   83     &   67800    &   93      \\
\hline
    \end{tabular}
  \end{center}
  \caption{
    Results of the fit to the ARROW simulations for the muon density
    at L3 level as a function of the distance from the shower core.}
  \label{muden}
\end{table}
%

\begin{figure}[htbp]
  \begin{center}
  \resizebox{0.9\textwidth}{0.52\textheight}{
  \includegraphics*{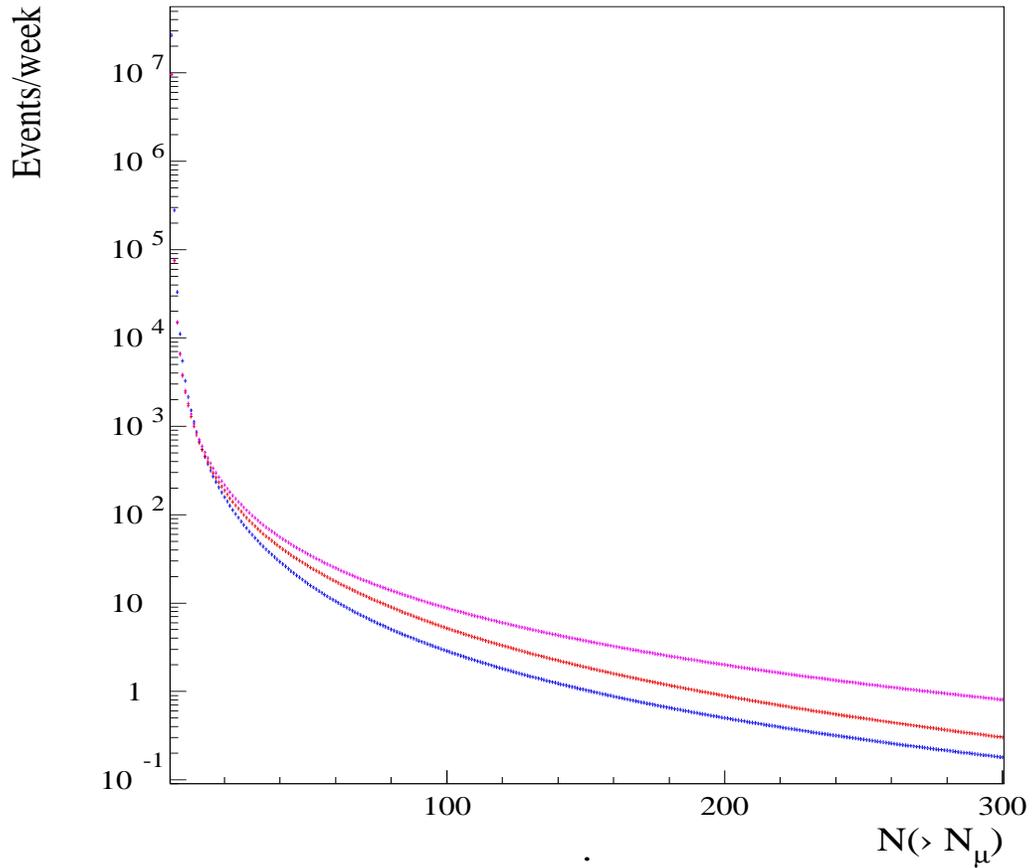}
  }
  \end{center}
  \caption{Integrated muon multiplicity - number of expected events with
           $\rm N(>N_{\mu})$ for a week of data taking.
           Lower curve - protons,
           middle curve - iron: case A,
           upper curve - iron: case B (see text).}
  \label{fig:mumult}
\end{figure}

Then a fast simulation of the expected muon multiplicity in the L3 detector
(suitable, after the necessary changes, for all LEP detectors) is 
developed. It is based on the fit results of the detailed simulations with ARROW,
using the parametrizations given above.
The muon multiplicity is assumed to originate from 
proton or iron induced vertical showers within 1 srad
with $\rm E_0$ from 1 TeV to 1 EeV.
The showers fall up to 1000 m from a detector with idealized 
geometry and \mbox{11 x 11 $m^2$} sensitive surface.
The primary flux is taken from~\cite{SH} in two limiting cases:
\begin{itemize}
 \item case A: the knee occurs at the same energy for each primary particle,
       leaving the composition unchanged in the whole energy range
 \item case B: the knee occurs at the same energy per nucleon,
       resulting in an increase of the iron component from 0.3 of the proton
       component below the knee to 1.5 well above the knee.
\end{itemize}

One should keep in mind that the parametrizations in~\cite{SH} contain
many components, so the flux used here amounts to 49.5 \% of the
total primary flux. This is good enough for the exploratory study presented 
in this talk, but certainly needs to be refined for a confrontation with the
real data.

The integrated muon multiplicity is shown in~\Figref{mumult}.
In one week we can expect to see events with more than hundred muons.
The distribution is very sensitive to the contribution of the heavy
component in the primary flux, as can be observed from the large
difference between cases A and B for iron nuclei.
A fast increase in the charged multiplicity with energy will also reflect
directly in this distribution.
In one year sufficient samples from showers above the knee can be accumulated.

%
%
\section*{Conclusions}

The results of this work can be summarized as follows:
  \begin{itemize}
    \item a Monte Carlo method, based on GEANT and GHEISHA, is developed
          for simulation of the hadron and muon components of EAS
    \item basic characteristics of muons at shallow depths underground
          can be computed in three dimensions
    \item the dependence of the muon flux on some key characteristics
          of the hadronic interaction model (inelasticity, multiplicity)
          is investigated
    \item a fast parametric simulation for the $\mu$ multiplicity underground
          is developed
    \item the $\mu$ multiplicity is a sensitive tool to study the 
          primary composition around the knee
    \item information about muon momenta and lateral spacing can give additional
          handles and will be included in more detailed studies
    \item the real challenge will be to compare the results of simulations with
          the data expected to be taken by the LEP detectors in the years
          1999 and 2000.
  \end{itemize}

%
%
\section*{Acknowledgements}
The author is grateful to P.~Le Coultre for the constant support of this
studies and to K.~Eggert for the invitation to participate in the
workshop.

%
%

\bibliographystyle{/l3/paper/biblio/l3stylem}
\bibliography{%
/l3/paper/biblio/l3pubs,%
/l3/paper/biblio/aleph,%
/l3/paper/biblio/delphi,%
/l3/paper/biblio/opal,%
/l3/paper/biblio/markii,%
/l3/paper/biblio/otherstuff,%
eth-pr-99-04}

\end{document}